\documentclass[fleqn,10pt]{wlscirep}
\usepackage[utf8]{inputenc}
\usepackage[T1]{fontenc}
\usepackage[normalem]{ulem}
\newcommand{\mox}{Mo$_8$O$_{23}$ }
\usepackage{url}

\title{Unveiling the electronic transformations in the semi-metallic correlated-electron transitional oxide Mo$_8$O$_{23}$}

\author[1*]{V. Nasretdinova}
\author[1,2]{Ya. A. Gerasimenko}
\author[2]{J. Mravlje}
\author[3]{G. Gatti}
\author[2]{P. Sutar}
\author[1,2]{D. Svetin}
\author[4]{A. Meden}
\author[2]{V. Kabanov}
\author[5,6]{A. Yu. Kuntsevich}
\author[3]{M. Grioni}
\author[1,2,7]{D. Mihailovic}

\affil[1]{Center of Excellence on Nanoscience and Nanotechnology Nanocenter (CENN Nanocenter), Jamova 39, 1000 Ljubljana, Slovenia}
\affil[2]{Jozef Stefan Institute, Jamova 39, 1000 Ljubljana, Slovenia}
\affil[3]{Institute of Physics, EPFL, Lausanne, Switzerland}
\affil[4]{Faculty of Chemistry and Chemical Technology, University of Ljubljana, Ve\v{c}na Pot 113, 1000 Ljubljana, Slovenia}
\affil[5]{P. N. Lebedev Physical Institute, 119991 Moscow, Russia}
\affil[6]{National Research University Higher School of Economics, Moscow 101000, Russia}
\affil[7]{Faculty of Mathematics and Physics, University of Ljubljana, Jadranska 19, 1000 Ljubljana, Slovenia}

\affil[*]{Venera.Nasretdinova@ijs.si}

\begin{abstract}

\mox is a low-dimensional chemically robust transition metal oxide coming from a prospective family of functional materials, MoO$_{3-x}$, ranging from a wide gap insulator $(x=0)$ to a metal $(x=1)$. The  large number of stoichometric compounds with intermediate $x$ have widely different properties. In \mox, an unusual charge density wave transition has been suggested to occur above room temperature, but its low temperature behaviour is particularly enigmatic. We present a comprehensive experimental study of the electronic structure associated with various ordering phenomena in this compound, complemented by theory. Density-functional theory (DFT) calculations reveal a cross-over from a semi-metal with vanishing band overlap to narrow-gap semiconductor behaviour with decreasing temperature. A buried Dirac crossing at the zone boundary is confirmed by angle-resolved photoemission spectroscopy (ARPES). Tunnelling spectroscopy (STS) reveals a gradual gap opening corresponding to a metal-to-insulator transition at 343K in resistivity, consistent with CDW formation and DFT results, but with large non-thermal smearing of the spectra implying strong carrier scattering. At low temperatures, the CDW picture is negated by the observation of a metallic Hall contribution, a non-trivial gap structure in STS below $\sim 170$K and ARPES spectra, that together represent evidence for the onset of the correlated state at $70$K and the rapid increase of gap size below $\sim 30$K.  The intricate interplay between electronic correlations and the presence of multiple narrow bands near the Fermi level set the stage for metastability and suggest suitability for memristor applications. 
\end{abstract}

\begin{document}

\flushbottom
\maketitle

\thispagestyle{empty}

\section*{Introduction}
The abundance of multivalent compounds, a malleable structure and the possibility of easy exfoliation has made molybdenum oxide MoO$_{3-x}$ family to be recognized early for battery applications\cite{Murphy79}. More recently they attracted  interest as prospective semiconductors for 2D electronics\cite{Yang19,patents} stemming from their high-$\kappa$ dielectric properties and high carrier mobility\cite{Balendhran13}.
The electronic properties of the metallic MoO$_2$ and semiconducting MoO$_3$ edge members among the molybdenum oxides are well understood within band theory\cite{Gopalakrishnan97}, and the band gap in the latter can be manipulated with a simple O reduction procedure\cite{Balendhran13}. Remarkably, removing a few MoO$_6$ octahedra from a unit cell or adding alkali atoms goes far beyond affecting the semiconductor properties but rather produces a wide range of stoichiometric suboxides Mo$_n$O$_{3n-1}$, the so-called ``Magneli'' phases. These phases have surprisingly complex phase diagrams, with richness reminiscent of that found in cuprates and other transitional metal oxides\cite{TMOreview}. The examples include, but are not limited to, the archetypal incommensurate CDW system, ``blue bronze'', K$_{0.3}$MoO$_3$, superconducting Li$_{0.9}$Mo$_6$O$_{17}$ \cite{Schlenker1,GreenblattOld} or the hidden nesting\cite{WhangboMo4O11,CanadellReview} CDW compounds $\eta-$Mo$_4$O$_{11}$ and KMo$_6$O$_{17}$. \mox stands out within the Mo$_n$O$_{3n-1}$ series for being a semi-metal with narrow bands, combining a small Fermi energy and low dimensionality that sets the perfect conditions for the emergence of strong correlations.

The complexity of \mox is seen already in the evolution of its properties with temperature, that cannot be ascribed to a single ordering process\cite{Nasretdinova2018}. The materials' properties are thought to be dominated by the high-temperature structural transitions that are usually associated with the onset of a CDW, but their nature is still far from understood. X-ray diffraction and neutron scattering studies~\cite{Sato,Fujishita,Fujishita87} reveal two structural transitions: the second-order-like one to an incommensurate state at $T_{CDW} \lesssim 360$, followed by an atypical lock-in transition at $T_{IC-C} = 285$\,K ($q_{\mathrm{IC}}=(0.195,0.5,0.120) \longrightarrow q_{\mathrm{C}}=(0,0.5,0)$). However, magnetic susceptibility\cite{Gruber83} is featureless at both transitions and resistivity\cite{Sato} exhibits weakly insulating behaviour below 300\,K, hence a gap opening is uncertain and the CDW scenario is not well established. Alternatively, the CDW origin of the observed structural distortion was challenged by the tight-binding ionic treatment of the band structure\cite{Whangbo}, suggesting that the soft mode of octahedral rotation is driving the transition instead, similar to the M$_3$ mode in perovskites. On the other hand, no noticeable softening around the structural transitions in either of the visible phonon modes were observed in the recent combined transient reflectivity spectroscopy and Raman studies\cite{Nasretdinova2018}, thus questioning this scenario too.

The lower temperature properties are even less understood, as several probes suggest further changes significantly below any reported structural transitions. At around $T^* \sim150$\,K, the resistivity experiences a pronounced maximum\cite{Sato}. At the same time, slightly above $T^*$, at $T\sim 160$\,K, time-domain electronic relaxation dynamics shows the appearance of a relaxation bottleneck\cite{Nasretdinova2018}, which was interpreted as an opening of a gap, that is unrelated to the CDW, on at least a part of a Fermi surface below this temperature.
Surprisingly, the most pronounced phonon mode also softens in this temperature range -- at $T_\mathrm{m} \sim 200$\,K. Its higher value, $T_m>T^*$,  could indicate competing orders\cite{Nasretdinova2018} and an incipient phase transition. At even lower temperatures, $T<30$\,K resistivity rapidly increases again, whereupon no changes were found by either structural or optical probes. A wide variety of temperatures at which anomalies are observed by different measurement techniques make the driving energy scale hard to determine and raise further questions whether one or more transitions occur below ($T_{IC-C}$).

To unveil the nature of the ground state and the sequence of transformations leading to it, here we explore the evolution of the electronic structure in the phase diagram of \mox using spectroscopic (STS, ARPES) and magnetotransport measurements on high-quality single crystals, and compare them with the DFT predictions. We find that despite quantitative agreement between DFT calculations and spectroscopic data on the scale of $\sim 1$\,eV, the low-temperature low-energy behaviour of \mox electronic spectra is that of a strongly correlated multiband material, in contrast with a narrow band semi-metal to semiconductor CDW-like transition model anticipated from DFT. This discrepancy is driven by the previously unresolved purely electronic transition at 70\,K with no structural involvement. The present data allows building of the unifying picture behind the phase diagram of \mox.

First, we elucidate the nature of high temperature CDW transitions. For the first time we observe the metal-insulator transition in resistivity at $T_\mathrm{CDW} = 343$\,K, accompanied by the mean-field-like opening of the partial gap of $2\Delta\sim 100$\,meV in STS spectra. The unusually large, non-thermal smearing shifts the onset of carrier depletion from $T_\mathrm{CDW}$ to below 200\,K, where the latter follows an Arrhenius law with a similar-sized gap. Moreover, the smearing results in a kink in Fermi level density of states at $T\sim T^*$, thus negating the presence of any phase transition at this temperature. We can consistently understand such behaviour through the tendency towards a high-temperature CDW formation due to the quasi-1D bands with the strong electron-hole asymmetry. The size of the gap $2\Delta\approx 100(50)$\,meV predicted by GGA (LDA) DFT is also consistent with the experimental observations down to 100\,K. However, this simple picture breaks down at lower temperatures.

Second, we identify the emergence of the new strongly-correlated ground state at low temperatures in the abscence of any reported structural transitions. The full gap of $\sim 130$\,meV, seen by all our probes, gradually develops below 70\,K. Below 70\,K Hall concentration first decreases and then saturates. The latter, together with the unusual spectral gap shape, suggests the nonuniform ground state with nanoscale phase separation, common for interacting systems. The signatures of interactions in narrow bands below the structural transitions set the stage for electronically driven metastability.

The paper is organized as follows. We first present confirmation that the studied surfaces indeed correspond to the pristine stoichiometric  \mox structure. We then move to presenting first-principle calculations of the band structure and compare them with ARPES and STS spectra at fixed temperatures. We then proceed by describing the transport and tunnelling spectroscopy studies in the full temperature range covering both high- and low-T transitions. Finally, we discuss all the data together in a self-consistent fashion in order to elucidate the unconventional behaviour of this material.

\section*{Results}
\begin{figure}[ht!]
\centering
\includegraphics[width=0.8\columnwidth]{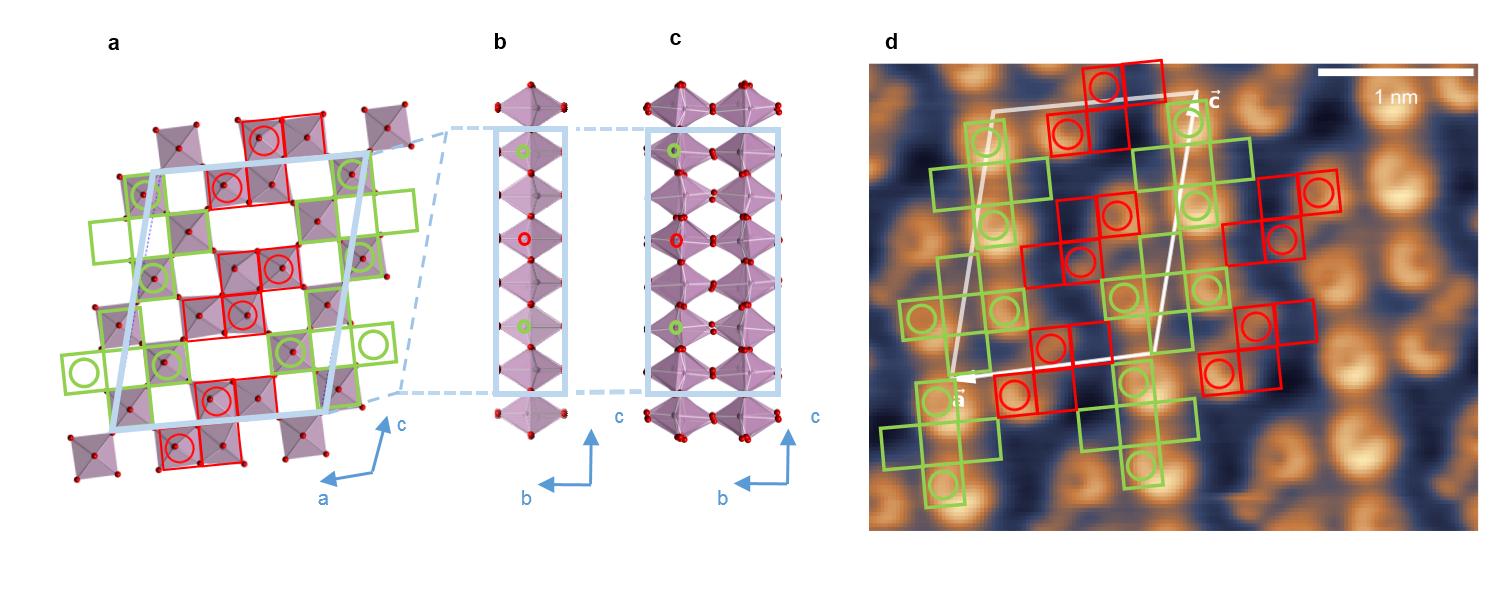}
\caption{\textbf{Structure and STM imaging of the \mox single crystal: }
(a-c) Crystal structure of \mox: (a) the high-T (010) ($ac$-)plane crystal structure (b) high-T $bc$-plane structure and (c) low-T $bc$-plane structure (based on crystallographic data from Refs.\cite{Sato,Fujishita}). Blue parallelogram
indicates the unit cell, empty circles show the Mo atoms that are shifted upwards from the equatorial oxygen plane of MoO$_6$ octahedron (that is discussed also in Refs.\cite{Fujishita,Whangbo}). (d) Pseudo-topographic STM image ($V_{tip} = +1.3$\,V; $I=100$\,pA) of (010) $ac-$plane at 4.2\,K. Thick white parallelogram shows the experimental unit cell. Green and red squares represent the assignment of
MoO$_6$ octahedra positions. Only those with Mo atoms shifted upwards (empty circles) from equatorial plane\cite{Fujishita,Whangbo} appear to be visible.}
\label{fig:rhostr}
\end{figure}
\subsection*{Surface structure}
Early STM investigations\cite{STMMo8O23} of \mox revealed two types of surface topographies, which were preliminary associated with the partial removal of an apical oxygen upon cleaving. While oxygen losses in the bulk are highly unlikely at the typical measurement conditions (see Methods), the surface changes might affect our results. To ensure that we study the properties of the pristine \mox surface, we start with claryfying the surface structure.

X-ray studies\cite{Magneli,Sato,Fujishita} show that \mox consists of layers of MoO$_6$ octahedra perpendicular to [010] axis, in which the Z-shaped clusters of Mo$_4$O$_{14}$ (shown in red on Fig.~\ref{fig:rhostr}a) are neighboured by the cross-shaped Mo$_4$O$_{15}$ clusters (shown in green on Fig.~\ref{fig:rhostr}a)\cite{Whangbo}. Additionally, the O-Mo-O bonds parallel to [010] axis are alternating in each MoO$_6$ octahedron\cite{Whangbo}, such that all the Mo atoms are always shifted either upwards or downwards from octahedra equatorial plane. The corresponding pattern for the high-temperature structure from\cite{Whangbo,Sato} is shown in the Fig.~\ref{fig:rhostr}a with the Mo atoms shifted upwards denoted by the large empty circles. Because of the opposite sense of Mo shift alternation in $ac$-plane along [001] $c-$axis (Fig~\ref{fig:rhostr}a,b), the unit cell of the high-temperature structure is (\mox)$_2$ and the space group is non-symmorphic with the $c$-glide symmetry between (\mox) subunits\cite{Whangbo}. The CDW distortion results in unit cell doubling along $b$-axis too as corresponding distortion and rotation of MoO$_6$ octahedra has the opposite character in a neigbouring layers; the bilayer unit cell of \mox after lock-in transition at $T_{IC-C} = 285$\,K is shown at Fig.~\ref{fig:rhostr}c.

Pseudo-topographic STM imaging ($V_{tip} = +1.3\,V$ and $I=100\,pA$, $T = 4.2\,K$) of the (010) plane reveals atomically flat surface with highly regular structure and the nm-sized unit cell (Fig.~\ref{fig:rhostr}d). The STM image appearance depends strongly on the bias, but stays the same in the $(0.6, 1.5)$\,V range, suggesting that it is the closest to a topographic one. The images taken at these biases remain the same also at elevated temperatures, at least up to $\sim200$\,K. The Fourier transform readily reveals the unit cell with the sizes very close to those seen by XRD. The surface structure assignment for this contrast is straightforward: only the octahedra with Mo atoms shifted upwards are seen in the image. They are shown with empty circles in Fig.~\ref{fig:rhostr}a,b,c. Unlike the previous studies\cite{STMMo8O23}, we observe the same surface structure in all the samples tested ($\sim$15) on the scale of $\sim$ 100~nm, thus ensuring that in ARPES and STS measurements we study the pristine \mox.

\subsection*{DFT electronic structure and comparison to STS and ARPES}
\begin{figure*}[htb!]
\includegraphics[width=1\textwidth]{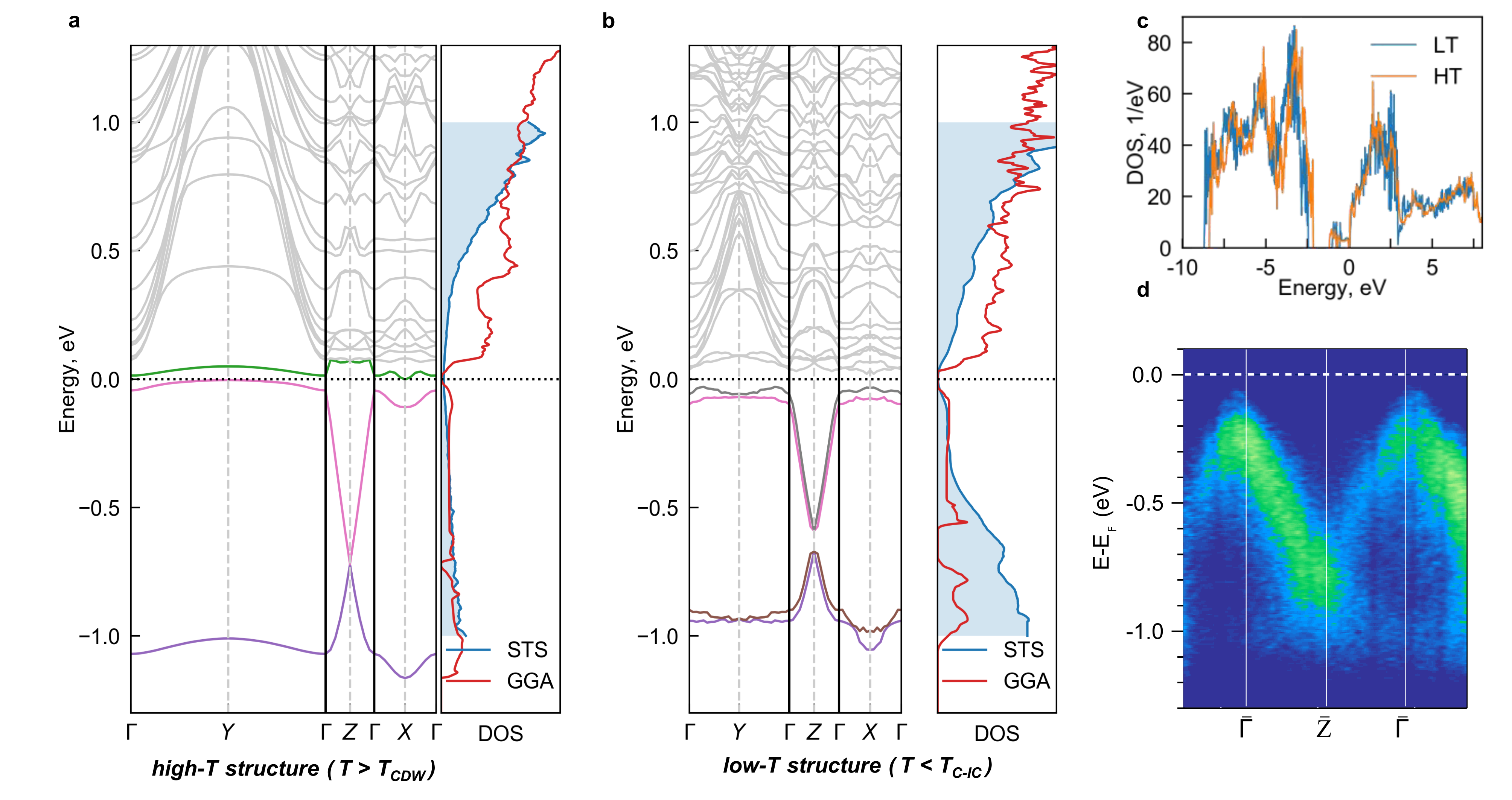}
\caption{\textbf{Comparison of calculated and measured band structures: }
(a) Calculated band structure for the high-temperature atomic structure $(T>T_{CDW})$. $\Gamma$ denotes the Brillouin zone centre, whereas $X= (0.5, 0, 0)$, $Y= (0, 0.5, 0)$ and $Z = (0, 0, 0.5)$ in the $(k_a=a^*, k_b=b^*, k_c=c^*)$ basis. Different colours correspond to a different bands. Right panel shows the comparison of DFT GGA density of states with the $dI/dV$ tunnelling spectrum at $T=403\,\mathrm{K}>T_\mathrm{ICDW}$ (setpoint: $1.3$\,V/1\,nA).  (b) Calculated band structure for the low-temperature atomic structure $(T<T_{C-IC})$. Right panel shows the comparison of DFT GGA density of states with the $dI/dV$ tunnelling spectrum at $T=4.2$\,K (setpoint: $1.3$\,V/1\,nA).  (c) Comparison of the low-$T$ and high-$T$ calculated density of states in the wide energy range.  (d) $\Gamma-Z$ band structure cut measured with ARPES at $T=20$\,K.
}
\label{fig:global}
\end{figure*}
\begin{figure}[ht!]
\centering
\includegraphics[width=0.9\columnwidth]{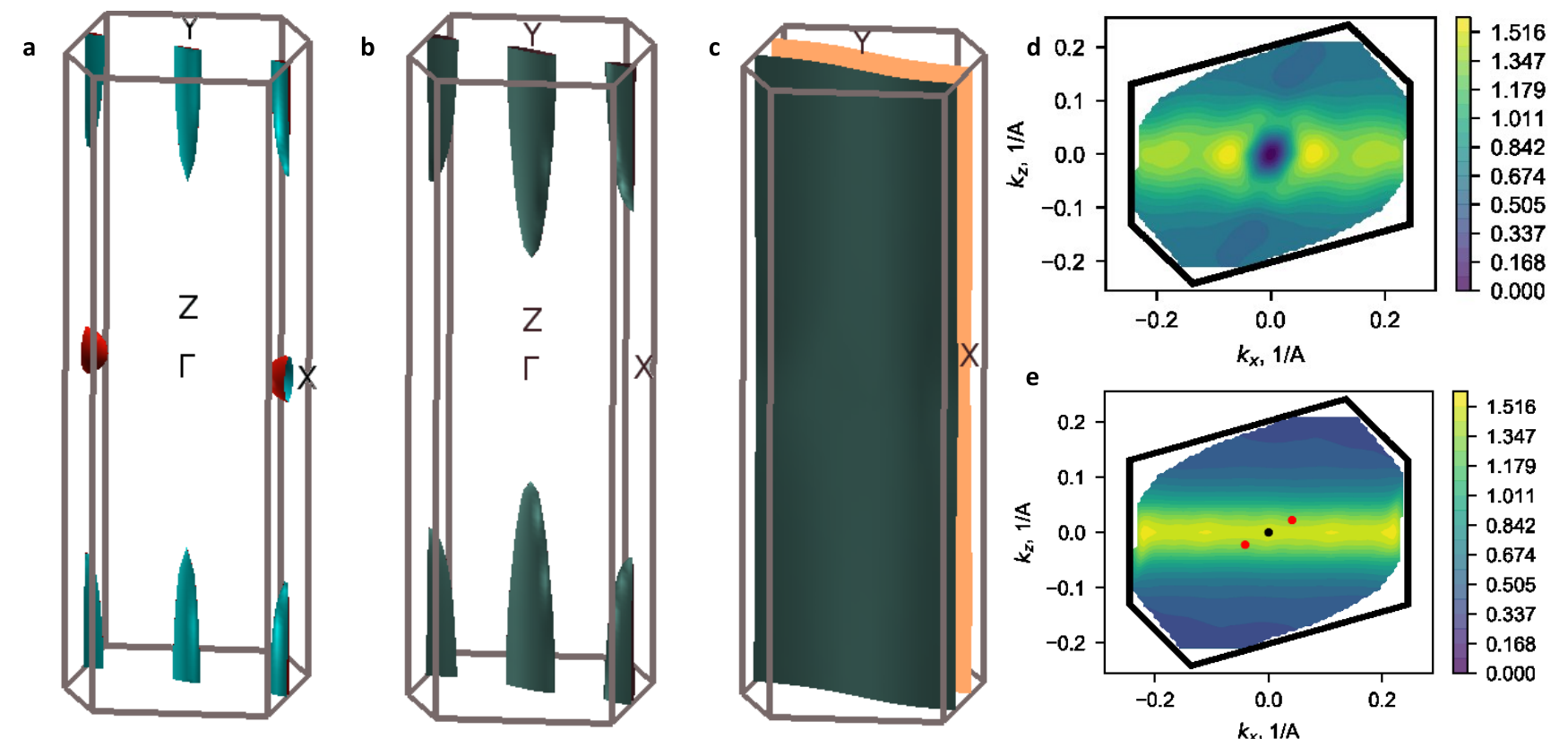}
\caption{
\\(a) Fermi surface for the high-T structure (LDA). Red --- electron pockets, blue --- hole pockets.
(b) equienergy surface evaluated at energy 50\,meV (c) and at energy 200\,meV below the top of the band. d) Lindhard susceptibility in the $k_x,0,k_z$ plane (e) Lindhard susceptibility in the $k_x,\pi/b,k_z$ plane, evaluated from the contributions of the two bands closest to the chemical potential at $T=350$\,K, retaining just band-diagonal contributions. Red dots are $q_\mathrm{ICDW}$ and -$q_\mathrm{ICDW}$ and black dot is $q_\mathrm{CCDW}$ according to \cite{Sato}}\label{fig:equidistantvsnesting}
\end{figure}

The generalized gradient approximation (GGA) DOS and bands for the high-temperature structure is shown in Fig.\ref{fig:global}a. The overall DOS shown at Fig.\ref{fig:global}c is very similar to MoO$_3$, with filled oxygen bands situated about $2$\,eV below the empty molybdenum bands. The main difference is the occurrence of the almost split-off lowest two molybdenum bands in the [$-1$\,eV, $0$\,eV] range. These two bands are (almost) fully occupied by the four electrons (note that the unit cell of \mox actually contains two \mox units, so the electronic states are in comparison with semiconducting MoO$_3$ filled with 4 more electrons). These bands demonstrate the characteristic Dirac crossings at $Z$ point due to non-symmorphic structure. The crossings can be also obtained from the group-theoretical treatment (see SI). The local-density approximation (LDA) band-structure (see Fig.~S1) is very similar, with the exception that valence bands and conduction bands are closer. The corresponding LDA Fermi surface sheets are displayed in Fig.~\ref{fig:equidistantvsnesting}a
with quasi-1D hole pockets near top and bottom of the Brillouin zone and ellipsoidal electron pockets in at $k_y=0$.

The GGA results for the low-temperature structure are shown in Fig.~\ref{fig:global}b. The unit cell is doubled in the $b$ direction, and hence one observes pairs of band mostly weakly influenced by the low temperature distortion with respect to the high temperature structure result. The main influence of the distortions is to fully open a (direct) gap that however remains small, with the bottom/top of the bands at $X$ point at $\pm 25$\,meV with respect to the chemical potential at $T=0$.

Prior tight-binding band structure calculations\cite{Whangbo} agree well with the present GGA results: the former have already captured qualitatively the gross features of \mox such as quasi-one-dimensional bands below the Fermi level and the semimetalic character of room-temperature properties. The GGA results allow now the quantitative assessment and direct comparison to the ARPES and STS data.

Tunnelling DOS in the $\pm1$\,eV range is roughly consistent with the DOS found in the GGA calculations. We compare the spectra obtained at $T = 4.2$\,K and at $T = 400\,\mathrm{K}>T_{CDW}$ above CDW transition, corresponding to the commensurate and undistorted crystal structures respectively\cite{Sato,Fujishita}. The DOS is strongly asymmetric with much larger weight above $E_F$ in the whole temperature range, in accord with the calculations. A notable discrepancy occurs at $E = -0.7$\,eV where the GGA calculations predict a suppression and (for the low-temperature structure) a small gap (see Fig.~S2).

The ARPES results reveal quasi-1D valence states dispersing along the $c$ direction, as expected from the crystal structure, and in agreement with the DFT calculations. Figure~\ref{fig:global}d is an energy-wave vector ARPES intensity map measured at $T=20$\,K along the $\overline{\Gamma}$-$\overline{Z}$ direction of the surface Brillouin zone. The measured dispersions are in overall agreement with the calculated band structure. The main hole-like band feature disperses linearly upwards from the BZ boundary to a maximum at $\sim -0.1$\,eV  at the $\overline{\Gamma}$ point. A second weaker electron-like band with a minimum at $\sim -1.1$\,eV  at $\overline{\Gamma}$ gains intensity after crossing the zone boundary.  Neither of these spectral features is as sharp as expected for a simple quasiparticle band. The anomalous energy and momentum width is consistent with the band splitting predicted for the low-temperature phase in Fig.~\ref{fig:global}b, but the underlying bands cannot be clearly resolved. This suggests that each band is further broadened by electronic correlations and/or electron-phonon interactions. The ARPES data do not provide evidence for the predicted gap at the $\overline{Z}$ zone boundary around $\sim-0.7$\,eV and, in this respect, are consistent with the DOS measured by STS.

\begin{figure}[t!]
\centering
\includegraphics[width=0.55\columnwidth]{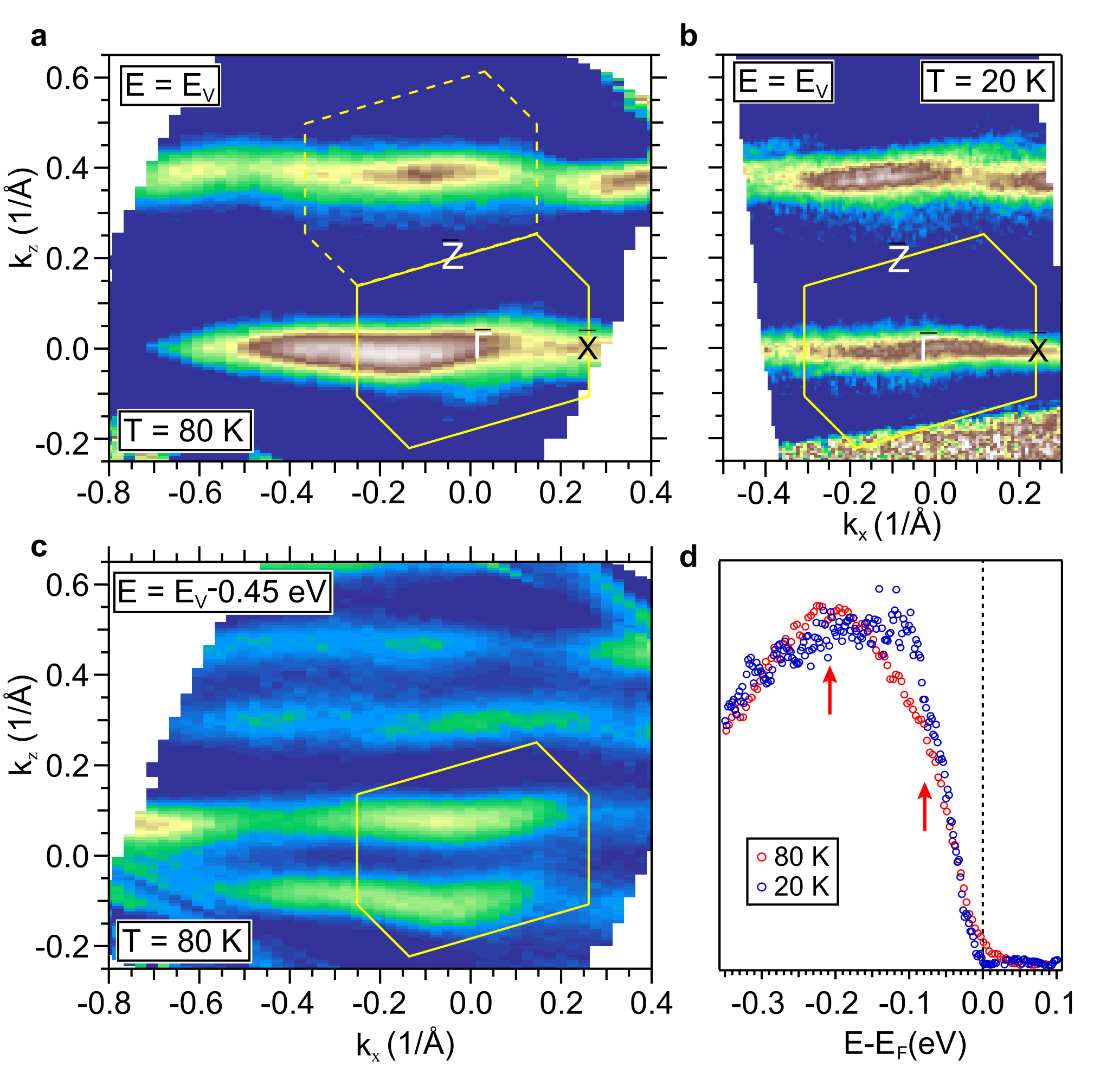}
\caption{\textbf{ARPES data:} a,b) Constant energy maps extracted at the top of the valence band E$_V$ at $T = 80$\,K and at $T = 20$\,K, respectively. The polygonal contour indicates the surface Brillouin zone. c) Constant energy map taken 450\,meV below $E_V$ at 80\,K. d) ARPES spectra measured at the $\Gamma$ point at $T = 20$\,K and at $T = 80$\,K.
}
\label{fig:arpes}
\end{figure}

The quasi-1D character of the bands is well illustrated by the ARPES constant energy maps in Fig.~\ref{fig:arpes}. The map in Fig.~\ref{fig:arpes}a, measured at the valence band maximum, shows parallel sheets aligned along $\overline{\Gamma}$-$\overline{X}$, i.e. perpendicular to the $c$ axis (the chain direction). These constant energy contours are split at larger binding energy, as shown in Fig.~\ref{fig:arpes}c, corresponding to the double intersections with the hole-like band on both sides of $\overline{\Gamma}$.  The parallel sheets are not perfectly straight. They exhibit wiggles with the periodicity of the BZ, which reflect the small inter-chain coupling. These wiggles are more clearly visible at 20\,K (Fig.~\ref{fig:arpes}b) where the contours are sharper. Similar wiggles are are also found in LDA equienergy surfaces.

Figure~\ref{fig:arpes}d shows spectra measured near the top of the valence band at $\overline{\Gamma}$ at 80\,K and 20\,K. Already at 80\,K the spectral weight at the Fermi level $E_F$ is very small, consistent with the presence of a gap at this temperature. In the broad line shape one can identify a peak at $-0.2$\,eV and a weaker shoulder at $-$0.07 eV, indicated by arrows, which can be associated with the energy splitting predicted by theory. We notice again that the energy width of the underlying features is not limited by the experimental resolution. At 20\,K the shallower feature grows in intensity and moves away from $E_F$ to $\sim -0.11$\,eV, suggesting a corresponding widening of the energy gap.

To understand where the tendency towards ICDW might come from, it is instructive to plot the equienergy surfaces in the momentum space. Fig. ~\ref{fig:equidistantvsnesting}b,c show that the occupied states have strongly 1D character
with dominant dispersion along the c*-direction (see also Fig. ~\ref{fig:global}); the character becomes even more pronounced deeper from Fermi level. To investigate the tendencies towards charge ordering instabilities, we evaluated the real part of the Lindhard electronic susceptibility (see Supplementary) that is shown in Fig.~\ref{fig:equidistantvsnesting}d,e. One indeed finds that the electronic susceptibilities are large and in particular finds the enhancement of the electronic susceptibility in a strip centred at $k_z=0$ with slightly higher values at $k_y=\pi/b$. This is in qualitative agreement with the experimental values, where the incommensurate ordering is found at $(0.195,0.5,0.120)$. The calculated susceptibility is large but not maximal at this wave vector (see SI for discussion).

\subsection*{Transport and Hall measurements}
\begin{figure}[htb]
\includegraphics[width=0.95\columnwidth]{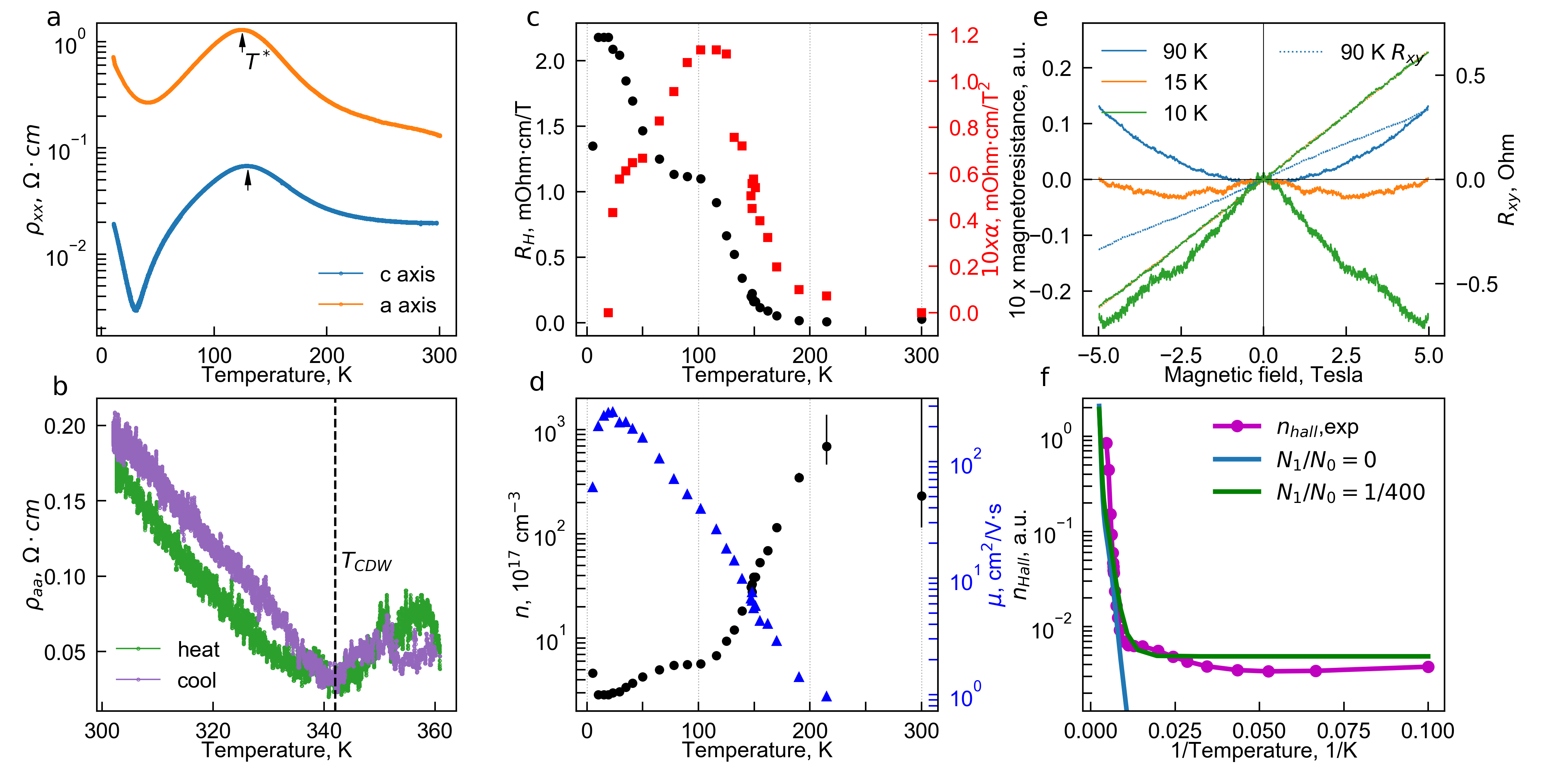}
\caption{\textbf{Transport, Hall effect and magnetoresistance measurements in Mo$_8$O$_{23}$ single crystal. }
(a) Temperature dependencies of the in-plane resistivity components $\rho_{xx}$ along $c$ and  $a$ crystal axes. Arrows indicates the temperature $T^*\approx 127$\,K of the resistivity peak. No apparent feature is found at the transition to the commensurate state, $T_{IC-C} = 285$\,K. (b) Inset shows high-temperature part of $\rho_a$ on heating and cooling. No history dependence or hysteresis is seen beyond the small deviations due to contacts degradation above 335\,K. The change of slope $d\rho/dT$ from metallic to insulating indicates the CDW onset at $T_{CDW} \approx 343$\,K.
(c) Temperature dependencies of Hall coefficient (black circles, left axis) and coefficient $\alpha$ in magnetoresistance (red squares, right axis). (d) Temperature dependencies of the Hall density (black circles, left axis) and Hall mobility (blue triangles, right axis) (e) Normalized magnetoresistance $\frac{\Delta R}{R} = \frac{R_{xx}(B)-R_{xx}(B=0)}{R_{xx}(B=0)}$ curves for various temperatures (left axis). Dotted curves show Hall resistance at corresponding temperatures (right axis).(f) Arrhenius plot of Hall concentration  (blue dots) and the two-band toy model (line). The model is given by $n_H \propto N_0*exp(-\Delta(T)/kT) + N_1$ where $N_0$ and $N_1$ correspond respectively to the DOS for the group b\#0 that is gapped at $T_\mathrm{ICDW}$ and group b\#1 that remains ungapped by that transition.}
\label{fig:Hall}
\end{figure}

The in-plane resistivity temperature dependencies, $\rho_{xx}(T)$, of \mox single crystals are shown in Fig.~\ref{fig:Hall}a. The overall temperature dependence is consistent with previously reported data\cite{Sato}, which was limited to below 300\,K. The in-plane anisotropy is $\rho_a/\rho_c \sim 5$ at room temperature, consistent with the one expected from band structure calculations. In the highest-temperature metallic state, just before any transitions, the resistivity along the most conducting $c$-axis is still rather high for a metal -- $\rho_c\approx20$\,mOhm$\cdot$cm, which, in the absence of significant disorder, points to large scattering and suggests the important role of correlations. It is also close to values expected for a polar metal\cite{Anderson65,NatMat2013}, which is interesting since \mox in CDW phase also lacks inversion symmetry\cite{Fujishita}.

Above room temperature, on cooling below $T_{CDW}\approx343$\,K, resistivity changes its slope from metallic to insulating (Fig.~\ref{fig:Hall}b), indicating the onset of the phase transition associated with the incommensurate charge density wave\cite{Sato}. Despite the broad resistivity feature at $T_{CDW}$, there is apparently no hysteresis at the transition temperature. The prehistory effects are limited to contact degradation which can be annealed. As temperature is lowered, resistivity increases with a minor change of slope at incommensurate to commensurate transition $T_{\mathrm{IC-C}}=285$\,K.

The most pronounced feature is the cusp at $T^*\approx130$\,K, below which the resistivity starts to drop. Below 30\,K the slope changes again to insulating. These features take place in the temperature range where no structural changes were reported. While they are observed in both in-plane components, the $T^*$ value is several Kelvin different between the two, emphasizing its link to transport properties rather than to the presence of some phase transition.

\label{mrandHall}
In a wide range of temperatures ($T>20$\,K), the Hall effect is linear-in-$B$, while the MR is quadratic-in-$B$ at least up to $\sim$10\,Tesla, see Fig.~\ref{fig:Hall}e. This is an indicator that the mobilities of the carriers involved in the transport are much lower than than 1/10 T$^{-1}\sim1000$\,cm$^2$/Vs. Indeed, maximum low-field Hall mobility estimate yields $\mu \approx 280$\,cm$^2$/Vs at $T=20$\,K. Hall effect shows n-type carriers over the whole temperature range. As temperature decreases from the room temperature, the resistivity, Hall effect and positive magnetoresistance (PMR) coefficient $\alpha$ ($ \rho_{xx}(B)=\rho_0+\alpha B^2$) -- all increase. We relate this growth with the decrease of number of carriers involved in conductivity with decreasing temperature.

In the $110-140$\,K interval, the resistivity and the $\alpha$ value have maxima, while the Hall resistance (Fig.~\ref{fig:Hall}c) has a plateau. Below $100$\,K the Hall concentration starts to decrease again, and the positive magnetoresistance drops steeply. The peak in PMR can be understood with the two-liquid model\cite{ZimanBook} as the result of matching conductivities of two group of carriers (see SI). Interestingly, these changes do not manifest themselves in the mobility (triangles in Fig.~\ref{fig:Hall}d): the latter grows monotonously as temperature decreases. Such behaviour suggests that only concentration is affected by the low-temperature transformation, whereas scattering mechanisms stay intact.

For the lowest temperatures (below $20$\,K), the negative magnetoresistance emerges, that becomes sharper as temperature decreases (Fig. \ref{fig:Hall}e). We relate the NMR to the weak localization phenomenon, as it is also accompanied by decrease of mobility.

\subsection*{Low-energy tunnelling spectroscopy}
\label{STSresults}
\begin{figure*}[t!]
\includegraphics[width=1\textwidth]{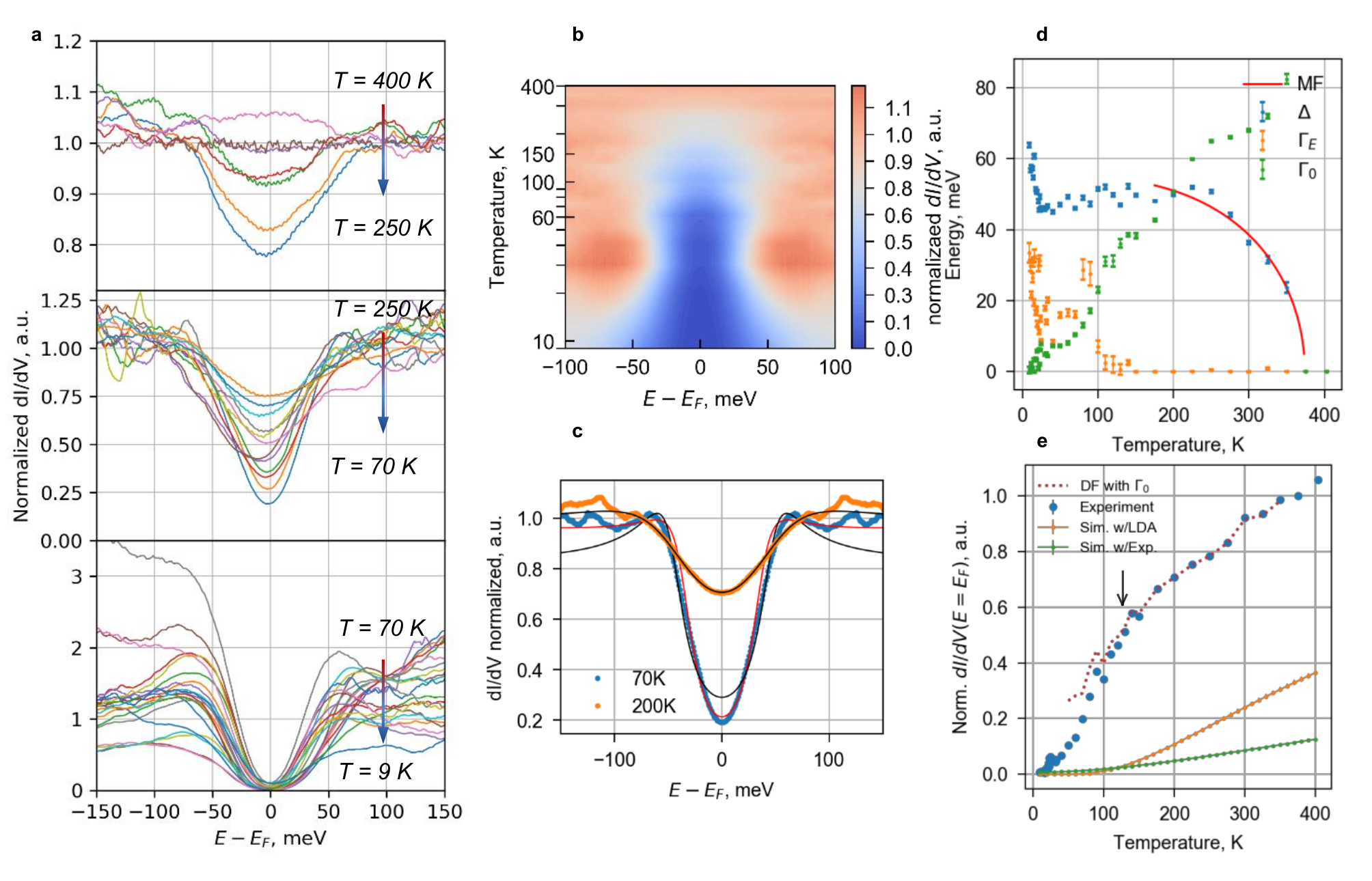}
\caption{\textbf{Tunnelling spectroscopy at low energies: }
a) Tunnelling conductance $dI/dV$ curves in the $9-400$\,K temperature range (normalized to the $dI/dV$ at $T = 375$\,K) and b) 2D plot of the same $dI/dV$ curves symmetrized for clarity
c) Examples of curves below and above $T = 175$\,K fitted with Dynes formula with constant $\Gamma$ (black) and energy-dependent $\Gamma(E)$ (red).
d) Temperature dependence of the gap value $2\Delta$, energy-independent smearing $\Gamma_0$ and energy-dependent $\Gamma_E$, extracted using modified Dynes formula from curves in (a). Red curve shows the $\Delta(T)$ fit for $T\in[176,350]$\,K with the mean-field (BCS) dependence.
e) Blue dots show the evolution of the normalized experimental Fermi level tunnelling density of states (raw $dI/dV$ values) with temperature. Orange and green curves show the theoretical Fermi level density of states evolution due to the thermal smearing of GGA and experimental 4.2\,K tunnelling spectra, respectively. Red dashed line shows the Fermi level density of states evolution for spectra obtained with the DF with $\Delta(T)$ and $\Gamma_0(T)$ substituted from the fits in (c) and the kink position is shown by an arrow.
}
\label{fig:sts}
\end{figure*}

The low-energy $\pm 100$\,meV tunnelling spectra reveal the complex two-stage gap opening from the initial semi-metal state as the temperature decreases from 400\,K to 9\,K (Fig.~\ref{fig:sts}a). The curves at 375\,K and 400\,K, above the CDW transition temperature, are featureless in this energy range and demonstrate almost constant finite tunnelling density of states with a shallow wide minimum at $E_F$. To emphasize the spectral changes upon cooling, we normalize all the curves by the 375\,K data, that is above the CDW transition temperature. Density of states starts to decrease in the $\pm50$\,meV energy range already below 350\,K. This spectral feature grows in width towards $\sim250$\,K. As temperature decreases, the dip becomes progressively deeper, reaching zero density of states at $E_F$ below 30\,K. At the same temperature, its width becomes larger (Fig.~\ref{fig:sts}b). This is preceded by the onset of the pronounced asymmetry in the spectra below 60\,K (lower panel in Fig.~\ref{fig:sts}a). On the other hand, no large-scale changes are observed around $T^*$, where transport crossover is observed.

Quantitative analysis of the $T\ge175$\,K tunnelling spectra with the Dynes formula\cite{Dynes78}(DF) reveals a mean-field-like gradual opening of the gap starting at $T = 350$\,K, which coincides with the MIT observed in resistivity. The DF satisfactorily describes the normalized spectra (Fig.~\ref{fig:sts}c) in the wide energy range, $\pm 0.3$\,eV, and allows us to extract independently the gap $\Delta$ and the phenomenological smearing factor $\Gamma_0$, shown in Fig.~\ref{fig:sts}d. The gap saturates around 250\,K, and further decrease in the density of states is due to decreasing smearing. Neither of the parameters experiences features around the $T_{IC-C}$ transition. Interestingly, the ratio of the saturation gap value and the transport MIT temperature, $\Delta_{sat}/kT_{CDW}\approx1.6$, is close to that for the weak coupling, quite unusual for inorganic CDW systems\cite{Monceau12}.

Simple DF analysis breaks down below $T\sim 175$\,K in estimating correctly the smearing, whereas the gap is reliably found to stay almost constant down to 40\,K whereupon it starts to grow again, increasing 1.5 times at 9\,K. The simple DF fit overestimates density of states at low energies, as shown in Fig.~\ref{fig:sts}c.

To describe consistently the spectra in the full temperature range, we introduce into DF the phenomenological dependence of smearing on energy (see SI), $\Gamma(E) = \Gamma_0 + \Gamma_E\left(\frac{E - E_F}{\Delta}\right)^2$. The choice of $\Gamma$ to be energy-dependent is motivated by it's high sensitivity to changing the fitting range in DF, quite unlike the gap $\Delta$ which is more robust. The form of $\Gamma(E)$ is given by the Taylor series expansion\cite{Dynes78} respecting the symmetry of the problem. With increasing the energy-dependent contribution $\Gamma_E$, the quasi-particle peaks at the gap edges are rapidly suppressed, while close to $E_F$ the spectrum is less affected, as illustrated for the 70\,K curve in Fig.~\ref{fig:sts}c. This corresponds to a significantly different shape of the tunnelling DOS, compared to the simple DF. We will use the ratio of $\gamma = \Gamma_E/\Gamma_0$ to quantify this difference.

The results of the fit are shown in Fig.~\ref{fig:sts}d. The spectrum starts to change when $\gamma$ becomes non-zero only below $T\sim150$\,K, reaching $\gamma\approx 1$ around 70\,K and rapidly peaking along with the increase of the gap size below 40\,K. In contrast, the usual smearing $\Gamma_0$ temperature dependence is slightly nonlinear and featureless. At high temperatures, it coincides with that in the conventional DF, reaching $\Gamma\approx2.7kT$, and approaches zero at low temperatures, along with zeroing of the density of states at $E_F$.

The observed large non-thermal smearing $\Gamma_0>\Delta$ strongly affects opening of the gap leading to the kink in the temperature dependence of the Fermi level density of states (FLDOS) close to $T^*$. Generally, FLDOS in a small-gap insulator should remain zero for $kT \ll \Delta$ and then smoothly increase with temperature due to thermal smearing. This general behaviour is indeed reproduced (Fig.~\ref{fig:sts}e) if low-temperature DOS is convoluted with the Fermi-Dirac distribution using the Bardeen formula for tunnelling~\cite{Chen} and is inconsistent with the data. For the kink to emerge, the temperature in simulations should be increased to $kT\sim\Delta$. The same effect is reproduced also with DF if $\Gamma_0$ becomes large compared to $\Delta$, as shown in Fig.~\ref{fig:sts}e for the experimental values of $\Delta$ and $\Gamma_0$.

Increase of the gap $\Delta$ and change of the DOS shape, $\gamma$, below 30\,K mark the stabilization of the new state at low temperatures, clearly different from the high-temperature one.
With $\Gamma_0\to0$ and finite $\Gamma_E$ at low temperatures, the tunnelling spectrum is completely different from the high-temperature DF shape. This change is clearly linked with the increase of the gap, implying that the latter is also unrelated to the order present at high temperatures. Linear extrapolation of $\Delta(T)$ gives the onset temperature of $\sim67$\,K, close to the point where $\gamma\approx1$, further supporting the close connection between the low-temperature state and change in the spectra shape. Importantly, the tunnelling spectrum at energies larger than the gap, $|E-E_F|>\Delta$, also starts to deviate from the high temperature one at 70\,K (bottom panel in Fig.~\ref{fig:sts}a), perhaps due to a change in tunnelling conditions or phase separation due to emergence of the new state.

To conclude, low-energy tunnelling spectroscopy provides us with strong evidence for the presence of two phases, one of which is related to a mean-field ordering transition, whereas another has correlated nature. The onset of the latter is clearly linked with the changes of background density of states at energies larger than the gap.

\section*{Discussion}
\textbf{High-T CDW transition}
For the temperature range $175-400$\,K, the overall results fit nicely the scenario of the semi-metal to incommensurate CDW phase transition. Indeed, for the first time we observe the metal-insulator like transition in resistivity at $T_\mathrm{ICDW}\approx343$\,K, (see Fig.~\ref{fig:Hall}b) together with the mean-field gap opening seen in the STS spectra (see Fig.\ref{fig:sts}a,d). The transition temperature is within the range expected from structural and neutron scattering studies $315$\,K$\leqslant T_\mathrm{ICDW} \leqslant 360$\,K\cite{Sato,Fujishita,Fujishita87}. The tendency to ICDW formation in \mox is lucid if one considers quasi-one dimensional character of the bands together with the temperature dependence of the chemical potential, that follows from the strong asymmetry of DOS of high-T band structure. The gap size predicted by DFT is consistent with the observations.

It is still puzzling why the Hall concentration follows the Arrhenius law expected for a gapped CDW only below 200\,K, well below $T_{ICDW}$. Similarly, why the soft mode observed in time-resolved reflectivity spectroscopy\cite{Nasretdinova2018} does not behave as expected for a CDW transition at $T_{ICDW}$, and softens only around 200\,K. Below we try to address these questions by looking into the spectroscopic signatures.

The combination of a large non-thermal smearing, $\Gamma_0/kT \sim 2.7$, and a rather small gap/transition temperature ratio, $\Delta/kT_{CDW}\sim 1.6$, observed both in STS and ARPES, leads to the regime of $\Gamma_0>\Delta$ at high temperatures, which makes \mox system clearly unconventional. For comparison, the ratio $\Delta/kT_{CDW}$ is several times larger in similar systems\cite{Monceau12}, and only slight deviation of smearing from thermal was reported for CDW in the same material class system Rb$_{0.3}$MoO$_3$\cite{Tanaka93}.
There are several contributions to smearing. Impurities and inhomogeneities of the order parameter can produce finite $\Gamma_0$, but its temperature dependence is weak in that case\cite{Hlubina16}. Inelastic phonon scattering\cite{Dynes78,Kaplan76} could be the reason for large $\Gamma_0$ values -- phonon peaks in Raman measurements\cite{Nasretdinova2018} are present up to $\sim100$\,meV, which is even larger than $\Gamma_0$ values seen in STS. The above mechanism was predicted to result in $\Gamma_0\propto T^3$, showing the tendency to saturation close to $T_C$ \cite{Mikhailovski91}. We indeed observe slight nonlinearity in energy-independent $\Gamma_0(T)$ above 30\,K suggesting that inelastic phonon scattering contributes substantially to smearing.

The regime of $\Gamma_0>\Delta$ changes the conventional carrier depletion usually observed in magnetic susceptibility or Hall measurements. Delocalized carriers contribute to Pauli and Landau terms in magnetic susceptibility -- both are proportional to the Fermi level density of states\cite{AbrikosovBook}. In conventional CDW systems it is expected to decrease abruptly after the transition, as it happens e.g. in  Mo$_4$O$_{11}$\cite{Gruber83}. However, in \mox magnetic susceptibility is featureless around $T_{CDW}$\cite{Gruber83} -- same as FLDOS extracted from STS -- and does not resolve the CDW transition. Instead, it changes slope at lower temperatures, where $\Gamma_0\lesssim \Delta$. Similarly, Hall concentration should follow the activation law after opening the gap, but in \mox it changes weakly down to $T\sim 200$\,K and only with further cooling the activation behaviour sets in (Fig.~\ref{fig:Hall}).
We thus conclude that the carrier density sees the gap only below this temperature, when $\Gamma_0$ becomes comparable to $\Delta$, despite the fact that the actual gap opens at much higher temperatures.

It is thus not surprising that the onset of the gap shape change $\gamma$ happens in the same temperature range where $\Gamma_0$ becomes comparable to $\Delta$. For the $\Gamma_0\gtrsim\Delta$ regime, small modifications to FLDOS, if present, are buried under the contribution arising from large smearing due to quasiparticles scattering. Once this contribution becomes small compared to the one produced by additional emerging order, the latter becomes visible in spectroscopic measurements. This is nicely illustrated by the same temperature for onset of $\gamma$ and Fermi level kink in Fig.~\ref{fig:sts}d,e. The kink in the Fermi level DOS at around 150\,K is most likely responsible for the transient reflectivity relaxation bottleneck and the increase of the optically-excited quasiparticles lifetime observed in the vicinity of this temperature\cite{Nasretdinova2018}. This extreme sensitivity of transient reflectivity measurements to the  delay in carrier depopulation suggests that anomalous softening of the most pronounced coherent mode close to 200\,K may also have the same origin. That would signify mode's polaronic nature, i.e. that of correlated electrons strongly interacting with the lattice. Interestingly, the polaronic nature of quasiparticles coupled to the high-energy phonons would be also consistent with the smearing of the photoemission spectra in other oxides\cite{GrioniTiO2}. Otherwise, the high-temperature bottleneck observed in the transient reflectivity relaxation\cite{Nasretdinova2018} below $T_\mathrm{CDW}$ agrees well with the present data: $2\Delta \sim 0.14$\,eV from the bottleneck fit is close to $2\Delta(T = 200 K)\sim 0.1$\,eV from STS.

\textbf{Low-T non-structural transformation.} 
In contrast to the high-T regime, the entire comprehensive set of data below $\sim 175$\,K directly confronts the simple phase diagram of a narrow-gap semiconductor or even that with only one CDW order parameter. Multiple experimental probes reveal that more than one carrier species is involved in the ordering processes which occurs more than 200\,K below the reported structural CDW transitions. Therefore, one has to conclude that the DFT band structure can describe well the gross features $\sim1$\,eV, but the low-energy DFT predictions, $< 0.1$\,eV, are inconsistent with the observations. We note that the narrow-gap semiconductor DFT scenario (see SI) still can be applied down to 100\,K to explain the resistivity maximum at $T^*$. Alternatively, DFT with a stronger influence from semi-metallic DOS can be considered within an excitonic insulator scenario and also used to explain the gapped behaviour (see SI).

The additional transformation takes place at $\sim70$\,K and it is not associated with the structural changes. The signatures are observed in multiple probes: (i)ARPES spectral shape changes between 80 and 20\,K, (ii) STS spectral shape changes around 90\,K, the gap size increases below 40\,K and can be extrapolated to 70\,K, (iii) Hall concentration decreases at 70\,K and then saturates at a different value. Below we will discuss the possible scenarios describing the above set of data.

Magnetotransport measurements yield the most direct evidences of at least two types of carriers: magnetoresistance peaks around $110-140$\,K and the Hall concentration demonstrates a plateau in the same range of temperatures. This behaviour can be approximated with a simple model $n_{Hall}^{2g} \propto N_0*\exp(-\Delta(T)/kT)+ N_1$ illustrated in Fig.~\ref{fig:Hall}f. Indeed, Hall concentration has a linear section in Arrhenius plot, giving a gap value of 100\,meV comparable to that in STS. The ratio of the high-T Hall concentration to that on the plateau is $N_0/N_1\approx400$, $N_1 \lesssim 10^{17}$\,cm$^{-3}$. Such behaviour of magnetotransport is expected when more than one band contributes to the Fermi surface: part of the Fermi surface is gapped due to CDW onset, whereas the small metallic pocket is left unaffected in a different part of the Brillouin zone.

The low-density group of carriers is further depleted below 70\,K in Hall measurements -- the same temperature where the correlation gap sets in the STS data.  Since we plot the $dI/dV$ curves normalized by the high-temperature spectrum, we do not resolve well the band occupied by $N_1$ carriers, only the change in it below 70\,K. The latter is likely related to a small kink around $50$\,K in FLDOS (Fig.~\ref{fig:sts}e) on the verge of our resolution. This depletion is also consistent with the magnetic susceptibility measurements\cite{Sato}, which revealed a small drop in $\chi_{deloc}$ (Pauli+Landau contributions of delocalized carriers) below $\sim70$\,K. Interestingly, no magnetic ordering so far has been observed\cite{Sato}, hence indicating that the change in $\chi_{deloc}$ is associated with the decrease of the FLDOS\cite{AbrikosovBook}.

Still, it is unusual that the change in a relatively small group of carriers affects the spectral shape on the energy scale of $2\Delta(T=0)\sim130$\,meV. Such a contribution to DOS can have two possible origins: spatial electronic inhomogeneity and orbital character (see SI). The data in this study was averaged spatially (see Methods). Therefore, in the case of a real space phase separation, the contribution to the total DOS coming from each phase is weighted by a relative area it occupies. Alternatively, in the case of reciprocal space phase separation, the contributions are weighted then by tunnelling selectivity -- the tunnelling matrix element. It is more selective to orbitals sticking out of the sample plane, compared to those in-plane\cite{Chen} (see SI).
The changes in tunnelling spectra below 70\,K thus allows us to conclude that the relative weight, either spatial or orbital, of the new state increases and is responsible for its enhanced visibility. At the same time, the increase of the gap size starts only below 40\,K, demonstrating that the increased gap size is directly related to the emerging low-temperature state, not the CDW.

Do the two states compete or coexist? We note that neither Hall concentration nor resistivity demonstrate activated behaviour at low temperatures, $T<70$\,K, in clear contrast to a simple band/CDW insulator, thus suggesting inhomogeneity or some additional contribution to transport. Macroscopic separation with large domains of either states can be ruled out because no hysteresis effects were observed in our resistivity measurements. Another scenario involves formation of static or dynamic nanoscale separation, which is quite generic in correlated systems with competing interactions\cite{Schmalian00}. The behaviour observed in \mox is thus reminiscent of that in the vicinity of the Widom line endpoint, characteristic of many Mott insulators. However, in the absence of reports on magnetic structure, dynamic coexistence of insulating and metallic nano-droplets, in the spirit of the Gor'kov-Sokol model\cite{GorkovSokol87}, appears the most favourable. Further real space spectroscopic imaging studies are necessary to address this issue.

To summarize, we have tracked the electronic transformations in the phase diagram of \mox, and found a new correlated-electron ground state, which emerges in the absence of known structural distortion. Using our new results we have proposed the robust unifying picture of various unexplained features previously observed in transport, magnetic and optical measurements. This work also brings a new light on the onset of the CDW in inorganic systems, emphasizing the role of electron lifetimes. The very large electron-hole asymmetry in the band structure implies that photoexcitation may lead to photodoping, and thus control of the phase diagram by light. Such perturbations could affect the transient electron occupation that would modify the preferred ordering wave vector. The competition of CDW and correlated state can be exploited for memristor devices or field-effect gating of in-plane resistance. 

\section*{Methods}
The samples were grown from Mo powder (99.9\% 1-2~$\mu$m), and Mo(VI) oxide taken in molar ratio 1:20 ($\approx 1$~g of mixture) thoroughly mixed and grind in agate mortar into a refined powder. Growth was done, similar to previous reports\cite{Magneli,Mo9O26growth}, with chemical vapour transport using iodine as agent. The crystals of \mox can be easily distinguished from other suboxides by pinkish color and their barrel shape. They can be nicely cleaved in the $ac$-plane. The monoclinic space group P2/c and unit cell parameters were confirmed at room temperature with the single-crystal X-ray diffraction (XRD) to match the previous reports\cite{Fujishita}. Thermogravimetric analysis demonstrated no change of the sample mass upon heating in Ar atmosphere up to 200\,$^\circ$C.

The resistivity tensor measurements were performed in vacuum on samples cut along corresponding crystal axis of approx. 1 mm $\times$ 300 $\mu$m x 100 $\mu$m with the conventional low-frequency ($7-14$\,Hz) AC four-point technique at the currents of $I\sim10-100\,\mu$A, in the linear $I-V$ regime. The samples were oriented using XRD. Magnetotransport measurements were done on a cleaved 180\,$\mu$m thick sample (lateral size $0.6\times0.7$\,mm$^2$) in van der Pauw geometry using Cryogenics dry CFMS-16 cryomagnetic system in the temperature range $2-320$\,K and magnetic fields up to $\pm10$\,T. The use of 50\,$\mu$A current ensured the linear $I-V$ regime throughout the whole temperature range. Due to intrinsic anisotropy of the sample and inevitably imperfect geometry, the diagonal component of the resistivity tensor admixtures to the Hall effect. High magnetic field range allowed us to extract magnetoresistance (symmetric in magnetic field) and the Hall signal (antisymmetric in magnetic field) measured simultaneously.

To achieve low contact resistance in (magneto)transport measurements, the pre-patterned contact areas were sputter-covered by the 50\,nm Au (40\,nm Ti for magnetotransport) layer. Samples were mounted by gluing thin annealed gold wires to the contact pads with the conducting silver epoxy and cured at 110$^\circ$C. To avoid strain, samples were left suspended on wires during the temperature dependence measurements. The wires were anchored to the sapphire substrate to ensure proper thermal contact.

All spectroscopic measurements were performed on fresh surfaces cleaved at room temperature in an ultra-high vacuum chamber, at a pressure of $\lesssim2\times10^{-10}$\,mbar. Tunnelling spectra and topography were measured with Pt tips; their spectra were checked on crystalline gold. Data at various temperatures were measured using the heated sample stage weakly coupled to the cryostat kept at either $4.2$\,K or $77$\,K. The temperature values shown are taken from the thermometer readings installed on the stage. Tunnelling spectra at each temperature were averaged spatially over several units cells taken from different sample spots within the area of $\lesssim200\times200$\,nm$^2$. ARPES measurements were performed on samples cooled to 80 K and 20 K, using 21.2 eV photons from of a high-brightness He lamp. The total spectral broadening (experimental and thermal) was lower than 80 meV for the data taken at 80 K and 20 meV for data taken at 20 K. The Fermi level was calibrated on a clean polycrystalline gold surface.

The density-functional theory calculations of electronic structure were done for previously identified crystal structures~\cite{Sato,Fujishita} using the Wien2k framework. The wave-vector summations were done over 500 and 1000 k-points in the irreducible Brillouin zone for the low-temperature and high-temperature structures, respectively. The energy cut off for valence electrons was set to $-6$ Rydberg, that is oxygen is described as He + valence electrons and molybdenum as argon + valence electrons. We performed the calculations both within the local-density approximation(LDA) and as well the generalized gradient approximation(GGA). Hubbard $U$ and spin-orbit coupling were not included in the present calculations. In related 4d correlated oxides, such as ruthenates or the rhodates, the spin-orbit coupling does not importantly affect the density of states.

\section*{Acknowledgements}
The authors acknowledge the financial support of Slovenian Research Agency (research core funding No-P1-0040, I0-0005 and No-P1-0175) and European Research Council
Advanced Grant TRAJECTORY (GA 320602), illuminating discussions with T.~Mertelj, and Evgeny Goreshnik for X-ray sample orientation. J.M. acknowledges support by Program P1-0044 of Slovenian Research Agency. G.G. and M.G. acknowledge the support of Swiss National Science Foundation. Magnetotransport measurements were performed using research equipment of the LPI Shared Facility Center. A.Yu. K. was supported by Basic research program of HSE.

\section*{Author contributions statement}
D.M. and V.N. conceived the project, P.S. synthesized the samples, A.M. performed the XRD characterization of the synthesized oxides, V.N., D.S. and Y.G. performed the transport measurements, V.N. and Y.G. performed the STM and STS measurements and analysed the results, A.Yu.K. conducted the magnetotransport measurements and analysed the results, G.G. and M.G. did the ARPES measurements and analysed the results. J.M. performed the DFT calculations and V.K. -- the group-theoretical analysis. All authors contributed to and reviewed the manuscript and supplementary information.

\section*{Additional information}
The authors declare no competing interests.

\end{document}